\documentclass[times, 10pt,twocolumn]{article} 
\usepackage{latex8}
\usepackage{times}

\usepackage{graphicx}
\usepackage{tabularx}

\begin{document}

\title{A Preliminary Work on Evolutionary Identification of Protein Variants and New Proteins on Grids}

\author{Jean-Charles Boisson, Laetitia Jourdan and El-Ghazali Talbi\\
LIFL/INRIA Futurs-Universit\'e de  Lille1\\ B\^at M3-Cit\'e Scientifique\\
\{boisson,jourdan,talbi\}@lifl.fr\\
\and 
Christian Rolando\\
Plateforme de Prot\'eomique / Centre Commun de Spectrom\'etrie de masse\\
\and
59655 Villeneuve d'Ascq Cedex, FRANCE\\
Christian.Rolando@univ-lille1.fr\\} 

\maketitle
\thispagestyle{empty}

\begin{abstract}
Protein identification is one of the major task of Proteomics researchers. Protein identification could be resumed by 
searching the best match between an experimental mass spectrum and proteins from a database. Nevertheless this approach can not 
be used to identify new proteins or protein variants. In this paper an evolutionary approach is proposed to discover new proteins 
or protein variants thanks a ``de novo sequencing'' method. This approach has been experimented on a specific grid called 
Grid5000 with simulated spectra and also real spectra. 
\end{abstract}

\Section{Introduction}
\label{introduction}

Proteomics can be defined as the global analysis of proteins. Protein identification is one of the major 
task of Proteomic researchers as it can help to understand the biological mechanisms in the living cells.  All the current
methods use data from mass-spectrometers and generally give good results. But in the case of protein variants or new 
proteins, these methods can only recognize a protein if it is stored in a database and can not clearly explain why this protein 
is different from any other in the database. The aim of our approach is to find the entire sequence of a protein, even in the case 
of variants or unknown proteins. To do that, we need to identify the different peptides that composed the protein. First, their
mass (their chemical formula) have to be found with a MS spectrum and secondly, from their mass, their sequence can be found
with MS/MS spectra. In fact, when peptides are known, we can obtain the complete protein. 

This article is organized as follows. Section~\ref{problem} deals with the specificities of protein variants and new protein 
identification problems; section~\ref{general_approach} describes our approach and the different algorithms that compose it; 
section~\ref{parallel} introduces the parallel framework; section~\ref{results} presents our results and discusses them and 
finally conclusions and perspectives about this work are provided.

\Section{The Positioning of the Protein Variants and New Proteins Identification Problem}
\label{problem}

The identification of new proteins and protein variants is a complex problem. All the existing protein identification 
methods are based on two types of data: MS and MS/MS spectra (MS for Mass Spectrometry) which are mass/intensity spectra. 
A MS spectrum is obtained by extraction of an experimental protein from a proteins mix, 
its digestion by a specific enzyme and its analysis in a mass spectrometer. From a MS spectrum, databases allow to identify all the 
peptides by their masses. Techniques using MS spectra for protein identification are identification methods by peptide mass fingerprint 
(PMF). The scoring of these methods is based of the comparison of an experimental peptide mass list with a theoretical
peptide mass list~\cite{AGMG1999,AMB2005}. They give good results but they only find the closest protein to the
experimental one without more information. A way to overcome the lacks of MS data is to use also MS/MS data (tandem mass spectrometry). 
Each peptide from the MS spectrum is selected and fragmented to obtain the corresponding MS/MS spectrum. The ions detected are 
characteristic of the structure of the parent peptide. Thus it is theoretically possible to obtain the sequence of each 
peptide from the digested protein. The use of MS data (mass of the peptides) combined to MS/MS data (partial sequence of the peptides)
 data increase the accuracy of the PMF techniques~\cite{ABE2001,AMMC2004}. These scores use several properties on the ions 
obtained by MS/MS spectra in order to find amino acid sequences. With partial amino acid sequences and masses, proteins
can be distinguished easier than with masses only. However, it is not sufficient to identify unknown proteins. 

An alternative method named \emph{de novo sequencing} has been proposed, using tandem mass spectrometry. It works on random sequence 
of proteins in order to find the experimental one (without databases). In this case the identification is based on random peptides 
or peptides result of a earlier identification (made by specific tools)~\cite{ADAC1999,AFP2005,AMHBJC,ASD2004}. But the MS/MS data are so 
fragmented (the deduced sequences are limited) and the number of theoretical protein that can be generated is so large that this 
kind of technique is only use on small amount of data. We speak about \emph{de novo \textbf{peptide} sequencing}. Furthermore, 
alignment tools as Blast are necessary to find the closest peptide corresponding to the result sequence and validate it.

Evolutionary approaches as optimization method have been already used against the huge research space of
the \emph{de novo peptide sequencing} problem~\cite{AHLR2004,AMHBJC} and give interesting results. So we have decided to design 
a genetic algorithm to make our \emph{de novo \textbf{protein} sequencing}.  

\Section{General Approach}
\label{general_approach}

According to the data available, the number of possible amino acid sequences is too huge to be enumerated.
So a genetic algorithm (GA) has been chosen for its ability to explore large solutions space. 

Find protein sequences needs two complementary steps: find the right peptidic masses with MS spectrum and from them
 find the corresponding sequences with MS/MS spectra.  The first step can be describe as follow: the individuals (randomly 
initialized) are digested (theoretical digestion) to be in a peptides list form and thanks to our evaluation function our GA 
can generate individuals that corresponding to the right peptidic masses list. We will now detailed each of these parts.

  \SubSection{Digestion Process}
  \label{digestion}

The digestion process corresponds to the cleavage of a protein in smaller residues called \emph{peptides}.
The cleavage points in the protein depend on the type of the used digestion enzyme because to each enzyme corresponds a
cleavage grammar. According to the chosen enzyme, the list of potential peptides is easily obtained. Nevertheless, 
in the real process, the enzyme can miss some cleavage points called \emph{miss cleavage}. So the number of potential peptides is
 greatly increased due to these miss cleavages. We developed a linear and iterative algorithm which realizes the theoretical 
digestion according to the grammar of the enzyme chosen and a number of miss cleavage allowed. Our algorithm works on a two-time 
basis: first, the peptides are computed  without miss cleavage and then, level by level the number of miss cleavage is
 increased until the wanted value.

The digestion process is an essential algorithm for Proteomics approaches. In the next paragraph, we will present 
the optimization method.

  \SubSection{The genetic algorithm (GA)}
  \label{ga}

\begin{figure}[h]
  \centering
  \includegraphics[width=8.25cm]{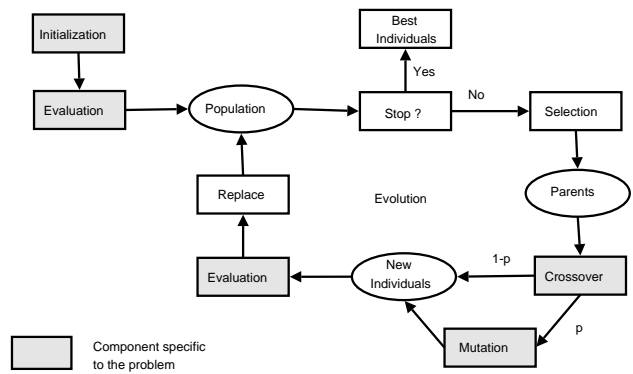}
  \caption{The general flowchart of a genetic algorithm. $p$ is the probability of mutation.} 
  \label{fig:aggen}
\end{figure}

  A Genetic Algorithm (GA) works by repeatedly modifying a population of artificial structures through the application of genetic 
operators (crossover and mutation)~\cite{hollan:ga}. The goal is to find the best possible solution or, at least good, solutions for 
the problem. Figure~\ref{fig:aggen} shows the global scheme of a genetic algorithm. Our GA has been developed thanks to 
the ParadisEO platform which is a C++ GPL (\textbf{G}eneral \textbf{P}ublic \textbf{L}icence) platform made for 
the conception of evolutionary algorithm~\cite{Cahon04a}. It may allow to find the right peptidic masses list corresponding to a 
MS spectrum.

\noindent

\begin{itemize}
\item{\textbf{Individuals Representation:} the chosen representation for an individual is a list
of peptides for 3 reasons: each individual is digested one time during the initialization process, the original sequence 
can be easily computed; the evaluation function and the fragmentation process need the proteins to be 
in a peptides list form. In details an individual is a list of peptides (with its number of miss cleavage), each peptide is an 
amino acid chain and each amino acid can have post-traductional modifications. 

}

\item{\textbf{Evaluation Function:} it is a completely original evaluation function based on a
optimized version of the algorithm developed by A.L. Rockwood~\cite{ARVO1995} to compute isotopic distributions.
The major interest of our function is a direct comparison of a experimental MS spectrum with a simulated one.
In fact our evaluation function does not need the mono-isotopic mass list extracted from the experimental MS spectrum. 
An individual of our GA is translated into a chemical formula list. 
For each chemical formula (so for each peptide), the isotopic distribution is gradually computed and, peptide by peptide, 
the simulated spectrum is calculated. The evaluation function computes the correlation between each theoretical peptide and
the experimental spectrum. So all the partial score of the theoretical peptides correspond to the fitness (the score) of
an individual. However, the evaluation function is time expensive: a protein of 500 amino acids needs one second 
in average to be evaluated.

This evaluation function has been validated by a research of known proteins in databases. To make our validation,
we use the UNIPROT database in FASTA format that can be download at \emph{www.expasy.uniprot.org/database/download.shtml}.
}

\item{\textbf{Individuals Initialization:} this process respects a \emph{de novo sequencing} approach. Individuals are
randomly generated according to a variable length (in amino acids). During the evolution of the GA, the size 
of individuals will change thanks to mutation operators (peptide insertion/deletion, amino acids insertion/deletion, amino
acid substitution and post-traductional modification mutations). A random generation allows to have a high 
diversity of population at the beginning of our search.
}

\item{\textbf{Operators:} they allow a diversified and intensified search. In a GA, there are two types of 
operators: the crossover operator and the mutation operator. The crossover operator allows to 
generate ``children'' individuals from ``parents'' individuals. In our case, we use the well known 1-point crossover 
operator. 

The mutation operator allows to have a genetic diversity in the new individuals. The individuals generated by crossover can have 
additional mutation. In our GA, there are 6 types of mutation: the random peptide insertion/deletion, the random
amino acid insertion/deletion, the amino acid substitution according a probability from a substitution matrix 
 (by default is the BLOSUM62 matrix~\cite{AHH1992}) and the post-traductional modification. The different mutations have an equal 
probability to be selected. All these operators allows the GA to get very close to the real biological model.
}
\end{itemize}

\Section{A parallel GA}
\label{parallel}
As we have previously noticed, the scoring function is time expensive. The GA was developed thanks 
ParadisEO~\cite{Cahon04a}. ParadisEO is one of the rare frameworks that provide the most common parallel and distributed models. 
These models concern the island-based running of metaheuristics, the evaluation of a population, and the evaluation of a single solution.
They are portable on distributed-memory machines and shared-memory multi-processors as they are implemented using standard libraries such
as MPI, PVM and PThreads. The models can be exploited in a transparent way, one has just to instantiate their associated ParadisEO 
components. 

\SubSection{Model}
As our scoring function is time consuming, we decide to parallelize the GA by simultaneously evaluating several
 individuals. The used model is a master/slave one. The master sends to slaves individuals to evaluate and the slaves send back
 the fitness value. The system is fault tolerant, the master can detect when a slave is available and send it a individual 
thanks a dispatcher. 

\SubSection{Infrastructure}

We decide to develop our project on a grid. Grid computing uses the resources of many separate computers connected by a
 network (usually the Internet) to solve large-scale computation problems. We use Grid5000 (\emph{www.grid5000.org})
 resources for our application. Grid5000 has resources located in Lille, Paris-Orsay, Rennes, Bordeaux, Toulouse, Lyon,
 Grenoble, Sophia Antipolis. Grid5000 uses Renater (the French national network for research and education) network to 
connect the different sites which speed is 2.5 Gbit/s.

\section{Results}
\label{results}

In this part, we will present the results of the GA and its parallelization. For all our experiments, the parameters of 
our GA have been set to 100 for the population size, 0.9 for the crossover rate (most used value) and 0.6 for the 
mutation rate (experimental value giving the best convergence speed). 

\SubSection{Biological validation}

In order to validate our first results, we compare the spectrum of our best individual with the simulated one of the
Apo-AI protein.

\noindent
\begin{figure}[h]
  \centering
  \includegraphics[width=8.25cm]{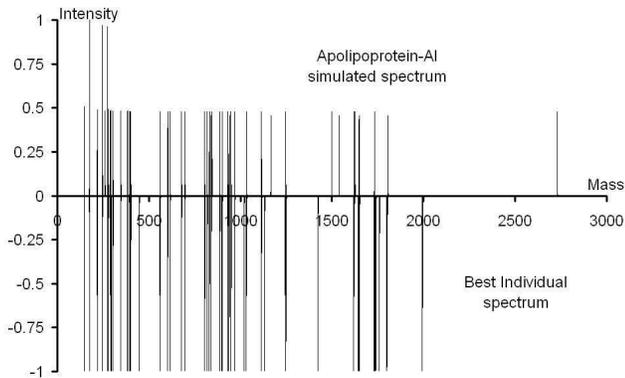}
  \caption{Apo-AI simulated spectrum vs best individual spectrum.}
  \label{fig:spectra}
\end{figure}

On figure~\ref{fig:spectra}, we see there a good correlation between the experimental spectrum and the best
individual of our GA. The most important value is the mass because, for the moment, all the simulated spectra that 
we generate have an intensity normalized to 1 (a high intensity only indicates that more than one peptide have the same mass).

\noindent
\begin{table}[h]
\centering
\begin{tabularx}{8.25cm}{|X|X|X||X|X|}
\hline
\multicolumn{3}{|c||}{Benchmark}& \multicolumn{2}{c|}{Best individual}\\
\hline
\centering{Proteins} & \centering{Type} & \centering{\# Peaks} & \centering{\# Peaks} & ~~~\# M\\
\hline
\centering{Apo-AI} & \centering{Sim} & \centering{43} & \centering{58} & ~~~~31\\
\hline
\centering{Apo-AI} & \centering{Exp} & \centering{20} & \centering{40} & ~~~~~9\\
\hline
\centering{Cyt-C} & \centering{Sim} & \centering{16} & \centering{15} & ~~~~~5\\
\hline
\centering{Cyt-C} & \centering{Exp} & \centering{26} & \centering{48} & ~~~~10\\
\hline
\centering{Albu}  & \centering{Sim} & \centering{65} & \centering{63} & ~~~~15\\
\hline
\end{tabularx}
\caption{Results gained for several type of data (Sim for simulated, Exp for experimental data, M for matches).}
\label{tab:perf}
\end{table}

Furthermore, Table~\ref{tab:perf} shows the results obtained with different types of data: simulated spectra computed from sequence 
in FASTA format and experimental spectra from mass spectrometer.
In this table, we remark that results on simulated data are better than results on experimental data. It's due to
convergence speed of the GA on simulated data. So we need to adapt the GA engine (precisely the number of generations 
and the stop criterion) to the specificities of the data. This can be possible when the second step of our approach will
be completely defined. 

\noindent
\begin{table}[h]
\centering
\begin{tabularx}{8.25cm}{|X|X|X|X|}
\hline
\centering{AAI pep} & \centering{$\delta$} & \centering{AAI pep} & ~~~~~$\delta$\\
\hline
278.153837 & 9.03 $10^{-5}$ & 839.339148 & 3.93 $10^{-5}$\\
\hline
347.229445 & 2.9 $10^{-3}$ & 886.474654 & 2.00 $10^{-5}$\\
\hline
381.213795 & 3.2 $10^{-5}$ & 899.441563 & 2.05 $10^{-5}$ \\
\hline
561.263264 & 7.19 $10^{-5}$ & 930.504892 & 1.07 $10^{-3}$\\
\hline
603.335367 & 1.64 $10^{-3}$ & 938.432714 & 9.94 $10^{-4}$\\
\hline
616.378235 & 1.7 $10^{-3}$ & 948.526690 & 1.18 $10^{-11}$\\
\hline
678.393885 & 1.5 $10^{-3}$ & 968.552905 & 6.25 $10^{-5}$\\
\hline
804.373937 & 6.03 $10^{-5}$ & 1114.585661 & 9.72 $10^{-4}$ \\
\hline
817.395676 & 2.27 $10^{-5}$ & 1247.576887 & 1.11 $10^{-5}$ \\
\hline
830.437206 & 1.24 $10^{-3}$ & 1647.801210 & 5.60 $10^{-6}$\\ 
\hline
\end{tabularx}
\caption{Matching Apo-AI peptides (AAI pep) and best individual peptides. $\delta$ is the mass difference, $\delta$= ($|$Apo-AI 
peptide - best individual peptide$|/$Apo-AI peptide). There are also 11 exact sequence matches which are not show here.}
\label{tab:mass_matching}
\end{table}

Table~\ref{tab:mass_matching} shows that the first of our approach is reached  because we find (globally) the right masses of peptide.
Although we have the correct chemical formula, we do not have necessary the right peptide sequence. But from the correct 
chemical formula and a MS/MS spectrum, we can extend the evolution of the GA to the right peptide sequence and so 
to the right protein sequence (second step for our approach). 

\SubSection{GA robustness}

A genetic algorithm is a stochastic algorithm and each execution does not always lead to the optimal solution.

\noindent
\begin{table}[h]
\centering
\begin{tabularx}{8.25cm}{|X||X|X|X|X|X|}
\hline
\centering{Data} & \centering{Max} & \centering{Best} & \centering{Mean} & \centering{Median} & ~~~~$\sigma$\\
\hline
\centering{AAI S} & \centering{186.54} & \centering{171.92} & \centering{163.19} & \centering{165.14} & ~~6.94\\
\hline
\centering{AAI E} & \centering{$\emptyset$} & \centering{63.15} &  \centering{61.30} & \centering{61.21} & ~~1.40\\
\hline
\centering{CC S} & \centering{35.09} & \centering{29.40} & \centering{23.75} & \centering{24.84} & ~~3.50\\
\hline
\centering{CC E} & \centering{$\emptyset$} & \centering{93.93} & \centering{88.07} &  \centering{88.31} & ~~4.13\\
\hline
\centering{AC S} & \centering{176.84} & \centering{170.56} & \centering{160.99} & \centering{165.26} & ~~9.23\\
\hline
\end{tabularx}
\caption{Statistics according to the data used. AAI: Apo-AI Human, CC: Cyt-C Bovin, AC: Albumin Chicken. S/E: simulated/experimental
 spectrum.~$\sigma$: standard deviation.}
\label{tab:stats}
\end{table}

  To study the behavior of the GA we perform 15 experiments (runs of the GA) for each protein. Table~\ref{tab:stats} summarizes 
some statistics over the experiments: the optimal fitness (in the case of experimental data, no protein matches exactly with the 
spectrum, so no value is given), fitness of the best individual, mean of the fitness solutions, median and standard deviation. 

Globally, our GA is quite robust on all the data as the median and the mean are very similar. We remark that we need to improve the GA 
to reach optimal value at each time.

\SubSection{Parallel version}
\label{parallel_version}

We experiment our parallel version on experimental proteins. We consider that $Ts$ is the time taken to run the fastest serial algorithm 
on one processor and $Tp$ is the time taken by a parallel algorithm on $N$ processors. To measure the gain of the parallelization, we 
compute two measures: the Speed-up = $S_N$= $\frac{Ts}{Tp} $ and the Efficiency = $\frac{S_N}{N}$.

\noindent
\begin{table}[h]
\centering
\begin{tabular}{|c|c|c|c|c|c|c|}
\hline
Nb & \multicolumn{3}{|c|}{Apo-AI} & \multicolumn{3}{|c|}{Cytc} \\
\cline{2-7}
Proc & Time & $S_N$ & Eff & Time & $S_N$ & Eff\\
\hline
1  & 5530 & 1 & 1 & 14712 & 1 & 1 \\
4  & 3430 & 1.61 & 0.4 & \textbf{3164} & \textbf{4.65} & \textbf{1.16}\\ 
8  & 1641 & 3.37 & 0.42 &  2055 & 7.16 & 0.9\\
16 & 1215 & 4.55 & 0.28 & 1443 & 10.2  & 0.64\\
32 &  759 & 7.29 & 0.22 & 1307 & 11.26 & 0.35\\
40 &  947 & 5.83 & 0.14 & 1020 & 14.42 & 0.36 \\
\hline
\end{tabular}
\caption{Execution time (in sec), Speed-up ($S_N$) and Efficiency (Eff) on Grid5000 for 2 experimental spectra 
according to the number of processors (Nb Proc).}
\label{tab:speed}
\end{table}

Table~\ref{tab:speed} summarizes values of the two measures for the Apo-AI and Cytc proteins.  We can observe that for 
Apo-AI the efficiency is less than 1 for any number of processors and is very bad for more than 32 processors whereas for Cytc
 we can observe supra linear performance. The reasons of such an observation can be due to load balancing (Grid5000 is an
 heterogeneous grid); a communication overhead or the potentially volatile nodes (Grid5000 is compound of PC clusters from 
university that could be potentially used by students or be turned off). 

\Section{Conclusions and Perspectives}
\label{conclusion}

In this article a genetic algorithm has been proposed to discover the sequence of an experimental protein. We have 
explained the limits of the current methods and the interest of a GA. The novelties of our approach are our 
evaluation function and the application of a \emph{de novo sequencing} method on complete proteins and not only on small peptides. 
Furthermore, we have experimented a parallel version of our GA on a Grid. A lot of work remains to increase 
the potential of the approach and the performance of the GA in order to find the right peptide sequences that compound the
experimental protein. We will continue to  work on our evaluation function in order to find new ones and manage to combine 
some of them in order to have better quality solutions. 

\bibliographystyle{latex8}
\bibliography{bibliography}

\end{document}